\pdfoutput=1
\documentclass[aps,prl,twocolumn, amsfonts, superscriptaddress]{revtex4}
\usepackage{mathptmx}
\usepackage{epsfig,amsopn}
\usepackage{graphicx}
\usepackage{amsmath,amssymb}
\usepackage{natbib}
\usepackage{braket}
\usepackage{color}
\usepackage{dcolumn}
\usepackage{bm}
\usepackage{times}
\usepackage{amsmath}
\usepackage{amssymb}
\newcommand\nn{\nonumber}
\newcommand\bea{\begin{eqnarray}}
\newcommand\eea{\end{eqnarray}}
\newcommand\be{\begin{equation}}
\newcommand\ee{\end{equation}}
\newcommand\f{\frac}
\newcommand\p{\partial}
\newcommand\la{\langle}
\newcommand\ra{\rangle}

\newcommand\ie{{\emph{i.e.}}}

\begin{document}

\title{Energy current cumulants in one-dimensional systems in equilibrium}

\author{Abhishek Dhar}
\affiliation{International  Centre for Theoretical Sciences, TIFR, Shivakote Village, Hesaraghatta Hobli, Bengaluru 560089, India}

\author{Keiji Saito}
\affiliation{Department of Physics, Keio University, Yokohama 223-8522, Japan} 

\author{ Anjan Roy}
\affiliation{ Max Planck Institute for  colloids and interfaces, Potsdam, Germany. }
\date{\today}

\begin{abstract} Recently a remarkable  connection has been proposed between the fluctuating hydrodynamic equations of a one-dimensional fluid and the Kardar-Parizi-Zhang (KPZ) equation for interface growth. This connection has been used to relate equilibrium correlation functions of the fluid to KPZ correlation functions.  Here we  use this connection to compute the exact cumulant generating function for energy current in the fluid system.  
This leads to exact expressions for all cumulants  and in particular to universal results for certain combinations of the cumulants. 
As examples, we consider two different systems which are expected to be in different universality classes, namely a hard particle gas with Hamiltonian dynamics 
and a harmonic chain with momentum conserving stochastic dynamics. Simulations
provide excellent confirmation of our theory.
\end{abstract}

\maketitle

The properties of current fluctuations of  energy and particle in various systems, both in and  out of equilibrium,  is an area of much activity recently. A number of papers have found unexpected universal features in these fluctuations in diverse systems \cite{bodineau04,eric13,derrida04,lee,saito07,derrida98,derrida99,appert08,brunet10,bertini15,roche05}. For example it has been shown that particle transfer in the symmetric exclusion process and charge transfer across disordered conductors 
have exactly the same value for a particular combination of current cumulants \cite{roche05}. On ring geometries \cite{brunet10} considered hard particle gases and looked at the net energy $q$ transferred across a section in a fixed large time interval $\tau$. Simulations indicated large fluctuations with the cumulants $\la q^2\ra_c/\tau \sim L^{-1/2}$ and $\la q^4 \ra_c/\tau \sim L^{1/2}$ where $L$ is the system size. This is in contrast to diffusive systems for which  $\la q^2\ra_c/\tau \sim L^{-1}$ and $\la q^{2n} \ra_c /\tau \sim L^{-2}$ for all $n$.

In this Letter we study energy current statistics of a system of interacting particles moving on a ring. We consider two different models, one deterministic and the other stochastic, but both having the same set of conservation laws. Our first model is the alternate mass hard particle gas\cite{garrido01,dhar02,grass02}  and the second model is the 
harmonic chain whose Hamiltonian dynamics is perturbed by an additional conservative noise \cite{BBO,leprietal09}. In both cases, energy, momentum and ``total stretch'' (see below) are conserved variables. These two models have been widely studied in the context of anomalous heat transport \cite{LLP03,dhar08} where they represent examples of  two  universality classes.  Recent work on fluctuating hydrodynamic theory (FHT) \cite{beijeren12,mendl13,spohn13,mendl14,das14} also predicts that these two models belong to different universality classes as far as the scaling form of various equilibrium  correlation functions is concerned. One of the 
main aims of this Letter is to use FHT to compute the cumulant generating function for current fluctuations (which is related to the corresponding large deviation function), look at universal features and their  differences for the two models, and compare the predictions with simulations.

We consider $N$ particles with positions and momenta described by the variables $\{z_x, p_x\}$, for $x=1,\ldots,N$, and moving on a periodic ring of size $L$ such that $z_{N+1}=z_1+L$ and $p_{N+1}=p_1$. The particles are assumed to only have   nearest neighbor interactions. In the alternate mass  hard particle gas (HPG), point particles move ballistically in between energy-momentum conserving collisions. The masses of the particles are chosen as $m_{2 x}=m_a,~m_{2x-1}=m_b$ for $x=1,2,\ldots,N/2$ (with $N$ chosen to be even).  The case $m_a=m_b$ is integrable while for $m_a \neq m_b$ one numerically observes ergodicity and equilibration (for $N >3$) \cite{garrido01,dhar02,grass02}, and it is expected that the system is non-integrable.  
The system is taken to be in equilibrium at time $t=0$ and we consider the statistics of the total heat transferred, $q(y,\tau)$, across a specified point $y$ in a given time interval $\tau$.  For the hard particle gas the energy flux at a spatial location $y$ is given by
\bea
j(y,t)=\sum_{x=1}^N \f{1}{2} m_x v_x(t)^3  \delta [y-z_x(t)]~.
\eea
The total energy flux in a fixed time interval is given by  
$q(y,\tau)=\int_0^\tau dt~j(y,t)$,  
and our interest is in  the statistics of this. On the ring geometry, 
the statistics is independent of $y$, hence we will omit the spatial index. Alternatively one can look at the statistics of the average integrated current, namely  
\bea
~~~Q(\tau)= \f{1}{L} \int_0^L dy~q(y,\tau)~. \label{Qeq}
\eea

We now briefly describe FHT for the HPG, and then apply it to compute the cumulant generating function for energy. In this theory, as detailed in \cite{spohn13}, one assumes that the conserved variables vary slowly in space and time. For the alternate mass gas, it is appropriate to consider a unit cell of size two and define the centre of mass variable $w_{x}=(m_a z_{2x-1}+m_b z_{2x})/(m_a+m_b)$, for $x=1,\ldots,N/2$. Next we consider the coarse-grained conserved fields $r(x,t)=w_{x+1}-w_x, p(x,t)=(p_{2x-1}+p_{2x})/2, e(x,t)=[p_{2x-1}^2/(2m_a)+p_{2x}^2/(2 m_b)]/2$. 
From the Hamiltonian equations of motion one finds 
\begin{align}
\f{\p r(x,t)}{\p t}&=\f{\p p(x,t)/\bar{m}}{\p x}, \nn \\
\f{\p p(x,t)}{\p t}&=-\f{\p {P}(x,t) }{\p x},\nn \\
\f{\p e(x,t)}{\p t}&=-\f{\p}{\p x}[ (p(x,t)/\bar{m}) {P}(x,t)]~, \label{eqm}
\end{align} 
where  $\bar{m}=(m_a+m_b)/2$, ${P}(x,t)=P(r,e)$,  is the local pressure and $\p f/\p x=f(x+1)-f(x)$ denotes the discrete derivative. 
The system is prepared in a state of thermal equilibrium at zero total average momentum, constant temperature ($T=\beta^{-1}$) and constant pressure ($P$) ensemble. This corresponds to an ensemble defined by the distribution 
$Prob(\{p(x),r(x)\})=\prod_{x=1}^{N/2} \frac{e^{-\beta [p(x)^2/2+ P r(x)]}}{Z(x)}~$, where 
$Z(x)=\int_{-\infty}^\infty dp \int_{-\infty}^\infty dr e^{-\beta [p^2/2+P r]}~.$ 
Now consider small fluctuations of the conserved quantities about their equilibrium values,
 $u_1(x,t)=r(x,t)- \la r \ra_{eq}$, $u_2(x,t) = p(x,t)$ and  $u_3(x,t)=e(x,t) -\la e \ra_{eq}$. 
The fluctuating hydrodynamic equations for the field $\vec{u}=(u_1,u_2,u_3)^T$ are now 
written by expanding the conserved currents in Eq.~(\ref{eqm}) to second order in the non-linearity and then adding dissipation and noise terms to ensure 
thermal equilibration. 
Thereby one arrives at the noisy hydrodynamic equations
\bea
\p_t u_\alpha= -\p_x  \left[ A_{\alpha \beta} u_\beta + H^\alpha_{\beta \gamma}  u_\beta u_\gamma - \p_x \widetilde{D}_{\alpha \beta} u_\beta + 
\widetilde{B}_{\alpha \beta} \xi_\beta \right]~.~~~~\label{EOM}
\eea
The noise and dissipation matrices, $\widetilde{B},\widetilde{D}$, are related by the
fluctuation-dissipation relation $\widetilde D C + C \widetilde{D} = \widetilde{B} \widetilde{B}^T$, where the matrix $C$ corresponds to equilibrium correlations and has elements  $C_{\alpha \beta}(x)= \la u_\alpha(x,0) u_\beta(0,0)\ra$. 
The noise term reflects that the dynamics is sufficiently chaotic, which indirectly rules out integrable systems.

We switch to normal modes of the linearized equations through the transformation $(\phi_{-},\phi_0,\phi_+)^T=\vec{\phi} = R \vec{u}$, where the matrix $R$ acts only on the component index and diagonalizes $A$, \ie~ 
$R A R^{-1}={\rm diag}(-c,0,c)$. The diagonal form implies that there are two 
sound modes, $\phi_\pm$, traveling at speed $c$ in opposite directions and one stationary but decaying heat mode, $\phi_0$.
The matrices $A$ and  $R^{-1}$ are given by
\begin{align}
 A &= \left(\begin{array}{ccc}
 0 & -1/\bar{m} & 0 \\
 \partial_\ell P &  0 & \partial_e P \\
0 & P/\bar{m} &  0 \end{array} \right)~, \nn \\ 
 R^{-1} &= \frac{1}{\sqrt{6} \beta P}\left(\begin{array}{ccc}
 -1 & 2 & -1  \\
-\sqrt{3 \beta\bar{m}} P &  0 & \sqrt{3 \beta\bar{m}} P \\
 P &  P & P \end{array} \right)~,~ \label{param}
\end{align}
where $\partial_\ell P$, $\partial_e P$ denote partial derivatives of the pressure with respect to average equilibrium stretch and energy respectively, and $c=(3 \beta/\bar{m})^{1/2} P$ is the speed of sound.
There is some freedom in $R$ and  it is  chosen  such that it satisfies the normalization condition $RCR^T=1$, where $C$ is the equilibrium correlation matrix \cite{spohn13}.

{\bf Current fluctuations}: 
The equations of motion Eq.~(\ref{EOM}), in terms of the normal modes, are given by
\bea
{\partial  \phi_s \over \partial t} &=&  -{\partial \over \partial x}\left[ j^N_s - D_s {\partial \phi_s  \over \partial x} + \sqrt{2 D_s} \xi_{s} \right]
\,
, ~~ s= -,0,+ \, , ~~ \label{eq1}
\eea
where $j^N_s = \sum_{\alpha} R_{s \alpha} j_\alpha$. 
Leaving out the fluctuation-dissipation parts, the various mode currents are given by 
\begin{align}
j^N_s&=c_s \phi_s + \sum_{s',s''}G^s_{s' s''} \phi_{s'} \phi_{s''}  \label{jnm}\\
{\rm where}~~ G^s &= \f{1}{2} \sum_{\alpha} R_{s \alpha} (R^{-1})^T H^\alpha R^{-1}~,~
c_{\pm }=\pm c,~c_0=0~. \nn 
\end{align}
The matrices $G^s$ are known completely in terms of microscopic parameters. Some of these for the HPG, that are relevant to our discussion here are: $G^0_{00}=0, 
G^+_{++}=-G^{-}_{--}=c/\sqrt{6},~G^+_{--}=-G^{-}_{++}=-c/\sqrt{6}$,~ $G^+_{00}=-G^{-}_{00}=0$ and $G^0_{++}=-G^0_{--}=c/\sqrt{6}$. 
Since the sound modes travel in opposite directions
while the heat mode stays fixed, we expect that the diagonal terms of $G^s$ in Eq.~(\ref{jnm}) are important for determining current correlations. Keeping only these terms, we get:  
\begin{eqnarray}
j^N_{\pm } &=& \pm c \phi_{\pm } + \f{c}{\sqrt{6}} (\phi^2_+ - \phi^2_{-})~, \nn \\
j^N_0 &=&  \f{c}{\sqrt{6}}  (\phi^2_+ -  \phi^2_{-})~. \label{jnmap} 
\end{eqnarray}
The energy current is given by $j_3=\sum_s \left[ R^{-1} \right]_{3 s} j^N_s =  (j^N_+ + j^N_{-} +j^N_0)/{\sqrt{6} \beta}$, and hence adding the contributions of the three modes we get
\begin{align}
j_3 &= \f{c}{\sqrt{6}\beta} (\phi_+  - \phi_{-} ) + \f{c}{2 \beta} (\phi^2_+ - \phi^2_{-})~. 
 \label{jen2} 
\end{align}
We note that the energy current  gets contributions {\emph{only from the sound modes}}.
Now we turn to the equations governing the three fields (\ref{eq1}). Again because of the fact that the fields quickly separate in space, it is sufficient to consider to retain only the leading nonlinear terms in the currents. Thus the sound modes are governed by  Burgers equations and this leads to the  correlations  satisfying KPZ scaling. On the other hand the heat mode which is diffusive at short times gets contributions from the sound modes and develops  Levy-like correlations at long times. 
The prediction of nonlinear fluctuating hydrodynamic theory \cite{spohn13},
 for the correlation functions of the three modes are
\begin{eqnarray} 
 \la \phi_\pm(x,t) \phi_\pm(0,0) \ra &=   \f{1}{(\lambda_s t)^{2/3}}~ f_{\mathrm{KPZ}} \left[~ \f{(x \pm ct)}{(\lambda_s t)^{2/3}}~\right]~\nn \\
 \la \phi_0(x,t) \phi_0(0,0) \ra &=  \f{1}{(\lambda_e t)^{3/5}} ~f^{5/3}_{\mathrm{LW}}\left[~ \f{x}{(\lambda_e t)^{3/5}}~\right]~,\label{eqscalE} 
\end{eqnarray}
where $f_{\mathrm{KPZ}}(x)$ is the KPZ scaling function, and $f_{\mathrm{LW}}^\nu(x)$ is the Fourier transform of the L\'{e}vy characteristic function $e^{-|k|^\nu}$.  The parameters $\lambda_{s,e}$ are known. 

The energy current in Eq.~(\ref{jen2}) is dominated by the sum of two counter-propagating Burgers modes $q=\int_0^\tau dt j_3 (t)= {P}/{(c\sqrt{2 \beta})}(q^N_+ + q^N_{-})$,  where $q^N_{\pm } =\int_0^\tau dt j^{N}_{\pm }(t)$. Hence, if $Z_{\rm BG}(\lambda) = \la e^{-\lambda q^N_+} \ra$  is the generating function for current in the Burgers equation, then the generating function of the energy current in the anharmonic chain is given by
\bea
Z(\lambda)= Z_{\rm BG}\left(\f{P}{c\sqrt{2 \beta}}\lambda\right)~ Z_{\rm BG}\left(-\f{P}{c\sqrt{2 \beta}}\lambda\right) ~. \nn
\eea
At large times we have $Z(\lambda)\sim e^{\mu(\lambda) \tau}$ and hence the energy current cumulant generating function $\mu(\lambda)$ is related to the Burgers current 
CGF $\mu_{\rm BG} (\lambda)$ as
\bea
\mu (\lambda)= \mu_{\rm BG} (\lambda)+ \mu_{\rm BG} (-\lambda)~.\label{cgf}
\eea
The CGF for the Burgers current is directly related to the current CGF in the ASEP. For ASEP on a lattice of size $N$ and particle density $\rho$, the CGF is known exactly and given by \cite{derrida98,derrida99}
\bea
\mu_{\rm{BG}}(\lambda) &=& \lambda \f{a^2}{2 \pi} +\f{a}{2 \pi N^{3/2}} G(\lambda a N^{1/2})\, ,  \label{cgfASEP} 
\eea
where $a = [2 \pi \rho (1-\rho)]^{1/2}$  and the function $G(z)$ is known through the following parametric equations
\bea
z = -\sum_{n=1}^\infty (-C)^n n^{-3/2} \,,~~
G(z) = -\sum_{n=1}^\infty (-C)^n n^{-5/2}~.\label{Gz}
\eea
The relation (\ref{cgf}) implies that the odd cumulants of the heat flux vanish while the even moments are given by:
\bea
C_{2 n}=\f{\la Q_\tau^{2n} \ra}{\tau} = 2 \left( \f{P}{c \sqrt{2 \beta}} \right)^{2n} c_{2n}~,
\eea
where the cumulants $c_{2n}$ are the  cumulants are obtained from a solution of 
Eqs.~(\ref{cgfASEP},\ref{Gz}) and given by
\begin{align}
C_2&=\f{a^3}{4 \pi 2^{1/2}  N^{1/2}}~, ~~
C_4=\left[~\f{9}{4} + \f{15}{4~ 2^{1/2}} - 2~6^{1/2}~\right] \f{a^5 N^{1/2}}{2 \pi}, \nn \\
C_6&=\left[~\f{1575}{8} + \f{8435}{24~2^{1/2}} - 50~3^{1/2} - 100~6^{1/2} - 36~10^{1/2}~\right]\f{a^7 N^{3/2}}{2 \pi}~. \nn
\end{align}
We can thus make the following predictions for the cumulants of the heat current in the mechanical HPG model:
(i) The $2$nd, $4$th and $6$th cumulants scale with system size as $N^{-1/2},~N^{1/2}$ and $N^{3/2}$ respectively.
(ii) For large system size we get the universal value for the ratio  
$r= {C_2 ~C_6}/{C_4^2} = 2.99248...$

We now turn to the second model, namely the harmonic chain with energy-momentum conserving stochastic dynamics. In the framework of FHT this corresponds to the special case of a symmetric potential and zero pressure, and expected to have different universality properties.  
The predictions of FHT is that the three modes satisfy the following equations
\begin{eqnarray}
\f{\p \phi_{\pm } }{\p t}
 &=& 
- {\partial \over \partial x} 
\left[ 
\pm c \phi_{\pm }  -D \f{\p \phi_{\pm }}{\p x} + \sqrt{2 D} \xi_{\pm }
\right] ~, \nn \\
\f{\p \phi_{0}}{\p t} &=& -\f{\p}{\p x} \left[ \left[G^{0}_{++} (\phi^2_+-\phi^2_{-})\right]- D_0 \f{\p \phi_{0}}{\p x} + \sqrt{2 D_0} \xi_{0} \right] \,  . ~~~~~~~
\end{eqnarray}
Thus the sound modes are diffusive while the heat mode can be shown to become Levy-like (with a different exponent than the $P \neq 0$ case). 

The energy current is now given by
\be
j_3 =  - \f{\kappa \p_\ell P } {c\sqrt{2 \beta}} G^0_{++}(\phi^2_+ - \phi^2_{-}). 
 \label{jen3}
\ee
Since the equations for the sound modes are linear, it is possible to obtain exactly the statistics of the energy current. Let us consider the characteristic function $Z_\phi(\lambda)=\la e^{-\lambda W} \ra$ 
where $ W=({1}/{N}) \sum_{l=1}^N \int_0^\tau dt \phi_l^2 (t)$ and the average is over the noise $\widetilde{\xi} (x,t)$. This involves Gaussian integrations and hence calculation is straightforward. However, since it is lengthy, the details are shown in supplementary material \cite{suppl}. We finally get $Z_\phi(\lambda)\sim e^{\mu_\phi (\lambda) \tau}$ with the cumulant generating function
\be
\mu_\phi(\lambda)= -\f{1}{2 \pi} \sum_{q \neq 0} \int_{-\infty}^\infty d \omega \log \left[ 1+
\f{\lambda}{N} \f{2 D \lambda_q}{\omega^2+D^2 \lambda_q^2} \right]~,
\ee
where $\lambda_q=q^2$ with $q=2s \pi/N$ and $s=1,2,\ldots$. Expanding $\mu_\phi(\lambda)$ in a series about $\lambda=0$ we get
\be
\la W^{n} \ra_c/\tau = N^{n-2} \f{(-1)^n B_{2(n-1)} }{ (n-1)! (2D)^n }~,
\ee
where $B_{2 n}$ are the Bernoulli numbers given by
\be
B_{2n} =\f{2 (-1)^{(n+1)}(2n)!}{(2 \pi)^{2n}} \sum_{s=1}^\infty \f{1}{s^{2n}}~.\nn 
\ee  
For the heat current corresponding to Eq.~(\ref{jen3}) we then get for the even cumulants:
\be 
C_{2n}=\f{\la Q^{2n}\ra_c}{\tau} = 2 A^{2n} N^{2n-2} \f{ B_{2(2n-1)} }{ (2n-1)! (2D)^{2n} }~,
\ee
where $A={\kappa \p_\ell P G^0_{++}}/{(c\sqrt{2 \beta})}$. In particular we get
\begin{align}
C_2 &= 2 A^2 \f{1/6}{(2D)^2}~, \nn \\
C_4 &= 2 A^4 N^2 \f{1/42}{ 3! (2D)^4}~, \nn \\
C_6 &= 2 A^6 N^4 \f{5/66}{5!(2D)^6} ~.
\end{align}
The $2$nd, $4$th and $6$th cumulants now scale as $N^0$, $N^2$ and $N^4$ respectively, which is completely different from the HPG scalings.  We again get the universal ratio $
r={C_2 C_6}/{C_4^2}={147}/{22}$.

{\bf Simulations}:
We now present results of simulations of  the two models, where we evaluate up to sixth cumulants, and compare with the predictions of the theory.   
For the case of the  hard-point gas, a system  of $N=L$ particles was taken, with  masses 
of alternate particles set at $1$ and $2.62$. This choice of mass ratio  is not crucial, and anything not too close to $1$ should work \cite{casati15}.   
The initial velocities of the particles are chosen from a microcanonical ensemble such that total momentum is zero and the total energy is $E=N$, which corresponds to $T = 2$. The initial positions of the particles are chosen from a uniform distribution between $0$ and $L$.  An event-driven molecular dynamics simulation was performed, in which successive  update time steps are taken to be the smallest of the collision times between  neighbouring pairs. 
After some initial transients, the system is run for a total time of $R \tau_{max}$ and we obtain $Q_r(\tau)$ for $r=1,2,\ldots,R$, and the sampling times are at regular intervals in the interval $(0,\tau_{max})$. The $n^{\rm th}$ moments are then computed as $\la Q^n(\tau) \ra=(1/R)\sum_{r=1}^R Q^n_r(\tau)$~. We then compute the cumulants.  The number of realizations averaged over was $R \sim {\cal{O}}(10^9)$.

\begin{figure}
  \includegraphics[width=3.2in]{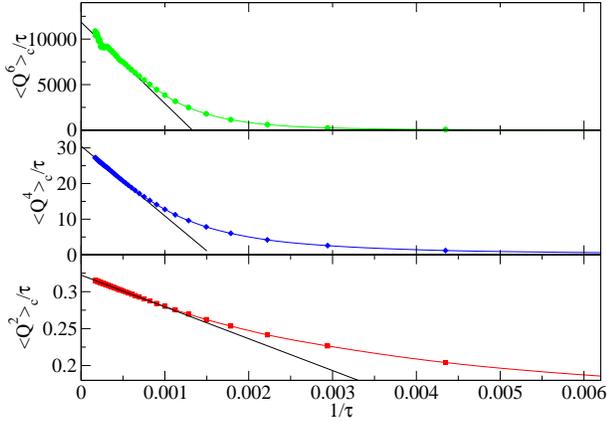}
\caption{(color online) Hard particle gas: Plot of $2^{\rm nd}$, $4^{\rm th}$ and $6^{\rm th}$ cumulants of integrated current divided by $\tau$ plotted against $1/\tau$ for a system of size $N=400$. The solid lines indicate the extrapolation procedure used to obtain the asymptotic value of the cumulant. }
\label{cumvst}
\end{figure}

\begin{figure}
  \includegraphics[width=3.2in]{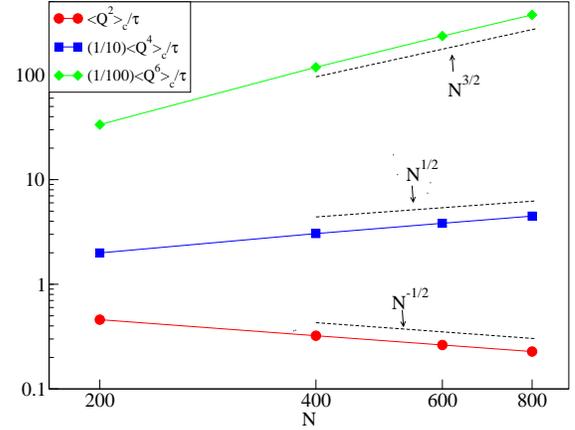}
\caption{(color online) Hard particle gas: Plot of $2$nd, $4$th and $6$th cumulants of the integrated energy current across a given point on a ring of size $L$ with $N=L$ particles, in the alternate mass hard particle gas. The higher cumulants are multiplied by constant numbers to make the plots clearer. The dashed  lines show the  expected slopes. }
\label{moms}
\end{figure}

 Cumulants obtained from the two definitions of heat flux, $Q$ in  Eq.~(\ref{Qeq}) or $q$, were evaluated. These behave differently at finite time, but both exhibit a linear growth for large $\tau$ and $\la Q^n\ra_c/\tau$ and  $\la q^n\ra_c/\tau$ appear to converge to the same value, as is expected. Here we show results only for $Q$. In Fig.~(\ref{cumvst}) we show the time dependence of the cumulants for a system of size $N=400$. Following \cite{brunet10}, we plot the ratios $\la Q^n_\tau\ra _c/\tau$  versus $1/\tau$ and by extrapolating the linear region of the graph, extract the asymptotic values. These asymptotic values of the cumulants have been plotted   in Fig.~(\ref{moms}). 
For $n=2,4$, our results agree with those of \cite{brunet10} while those for $n=6$ are new results. 
We find very good agreement with the predictions of the theory for  the system-size dependence of the cumulants.

\begin{figure}
  \includegraphics[width=3.2in]{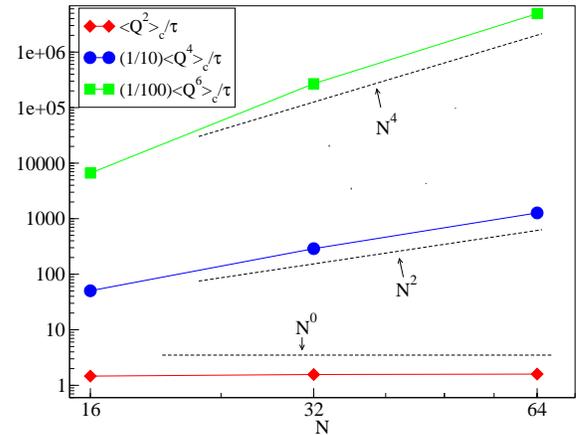}
\caption{(color online) Momentum exchange model: Plot of $2$nd, $4$th and $6$th cumulants of the integrated energy current across a given site on a ring with $N$ particles. The higher cumulants are multiplied by constant numbers to make the plots clearer. The dashed  lines show the  expected slopes.}
\label{excmoms}
\end{figure} 

For the momentum exchange model also, our simulations are done in the zero total momentum ensemble. In this case it appears numerically challenging to study very large system sizes, possibly because the cumulants become extremely large with 
increasing system size. Averages were done over $R \sim {\cal{O}}(10^7)$ realizations for sizes up to $N=64$. However, in this model, asymptotic results are 
known to be reproduced even at relatively small system sizes \cite{leprietal09}. 
The values of cumulants obtained by  linear extrapolation, have been plotted in  in Fig.~(\ref{excmoms}). Again we find a good agreement with the predictions of the theory for  the system-size dependence of the cumulants.

The value of the ratio $r$, a number constructed from quantities with very different order of magnitude, is found to have the same order of magnitude as the predicted value though the number is a bit different. 
Some of the predictions of FHT are expected to be valid at short time scales when heat and sound modes do not interfere. On the other hand, while computing current fluctuations, one is looking at finite systems and large times during which fluctuations travel many times around the ring. An assumption we make is therefore that the interaction between the heat and sound modes is weak. The deviations in the expected universal value of  ratio of $r$ could be attributed to this.

In summary, we have shown that the recently developed fluctuating hydrodynamics (FHT) for one dimensional systems can also be used to obtain predictions for energy current fluctuations  on the ring geometry. It is seen that the energy current statistics are determined by the statistics of sound mode fluctuations which satisfy the Burgers equation (for HPG gas) and the diffusion equation (for the  harmonic 
chain with noisy dynamics). Remarkably it is found that the system size-dependence of the cumulants is completely different for the two universality classes. Simulation results  support our analytic conclusions.  The predicted system size dependence is also consistent with the predictions obtained from a Levy walk model for anomalous transport \cite{dhar13} if one assumes that the relaxation time scale is determined by decay of sound mode fluctuations. 
FHT has recently found diverse applications and fascinating results have been obtained in systems such as the discrete nonlinear Schr\"odinger equation \cite{kulkarni2015} and  exclusion processes on coupled lattices \cite{popkov2014}. It will be of great interest to see if results on current statistics in these systems can also be obtained using the ideas proposed here.

\textit{Acknowledgments:} We are grateful to Bernard Derrida   for his many valuable suggestions and for a critical reading of the manuscript.  AD acknowledges support from  UGC-ISF Indo-Israeli research grant F. No. 6-8/2014(IC). KS was supported by JSPS (No. 26400404). We thank the Galileo Galilei Institute for Theoretical Physics for the hospitality and the INFN for partial support during the initiation of this work. We thank the NESP program at the International centre for theoretical sciences, TIFR, where this work was completed.

\newpage

\widetext
\bigskip
\noindent\rule{\hsize}{1.5pt}
\bigskip

\setlength{\baselineskip}{20pt}

\begin{center}
{\large \bf Supplementary Material for \protect \\ 
``Energy current cumulants in one-dimensional systems in equilibrium'' }
\end{center}
\vskip 1cm

We consider the energy fluctuation for zero pressure case. 
In this case the predictions of fluctuating hydrodynamics theory is that the
three modes satisfy the following equations
\begin{align}
\f{\p \phi_+}{\p t} &=-c \f{\p \phi_+}{\p x}  +D \f{\p^2 \phi_+}{\p x^2} + \f{\p}{\p x} \sqrt{2 D} \xi_+~, \nn  \\
\f{\p \phi_{0}}{\p t} &= -\f{\p}{\p x} \left[G^{0}_{++} (\phi^2_+-\phi^2_{-})\right]+ D_0 \f{\p^2 \phi_{0}}{\p x^2} + \f{\p}{\p x} \sqrt{2 D_0} \xi_{0} ~, \nn \\ 
\f{\p \phi_{-}}{\p t} &=c \f{\p \phi_{-}}{\p x} + D \f{\p^2 \phi_{-}}{\p x^2} + \f{\p}{\p x} \sqrt{2 D} \xi_{-} ~.  \label{P0mode-eq}
\end{align}
Thus the sound modes are diffusive, while the heat mode can be shown to become Levy-like (with a different exponent than the $P \neq 0$ case). 
The energy current is now given by
\be
j_3 =  - \f{\kappa \p_\ell P } {c\sqrt{2 \beta}} G^0_{++}(\phi^2_+ - \phi^2_{-}). 
 \label{jen3}
\ee
Hence, to obtain the cumulant generating function (CGF) for the energy current, it is sufficient to compute the CGFs for the
currents associated with the two independent diffusive fields  $\phi_{\pm}$. 
We now proceed to compute the CGF of a single diffusive field, say $\phi$. Let us consider the space-discretized, continuous time  diffusion equation given by
\bea
\p_t \phi_l(t) = D( \phi_{l+1}-2 \phi_l +\phi_{l-1} ) + B (\eta_{l+1}-\eta_l)~,
\label{eqmP0}
\eea
where $\phi_l(t)$ is the field variable of the sound mode at the $l$th site and at time $t$. 
In the calculation, the sound velocity does not contribute to the result, and hence we omitted this term at this stage. We dropped $\pm$ subscript in sound mode variables. The variable $\eta_l$ are taken to be  white Gaussian noise terms with unit variance, and their strength $B$ is fixed such that equal-time correlations at long times converge to the expected equilibrium value $\la \phi_l(t) \phi_m(t) \ra \to \la \phi_l \phi_m \ra_{eq}=\delta_{l,m}$.

Let us define the Fourier transforms and inverses by:
\bea
&&\{\phi_l (t),\eta_l(t)\} = \sum_q \sum_w \{\widetilde \phi (q,w), \widetilde \eta (q,w)\} e^{i (q l-\omega t)} \nn \\
&&\{ \widetilde{\phi} (q,w), \widetilde \eta (q,w)\} = \f{1}{N \tau} \sum_{l=1}^N \int_0^\tau dt \{\phi_l (t), \eta_l (t)\} e^{-i (q l- \omega t)} \nn, \\
&&{\rm where}~~q=2 s \pi/N,~ w=2 n \pi/\tau~.  
\eea
Plugging this into Eq.~(\ref{eqmP0}) we get
\bea
\widetilde{\phi} (q,\omega) = \f{B (e^{i q}-1) \widetilde{\eta} (q,\omega)} {-i \omega +D \lambda_q}~,~~~ {\rm where}~~\lambda_q&=2 [1-\cos (q)]~.
\eea
The noise correlations are given by $\la \widetilde{\eta} (q,\omega)\widetilde{\eta} (q',\omega')  \ra = [1/(N \tau)] ~\delta_{q+q'} \delta_{\omega+\omega'}$. This leads to the following expression for equal time correlations:
\bea
\la \phi_l(t) \phi_m (t) \ra = \f{1}{N \tau} \sum_q \sum_{\omega} \f{B^2 \lambda_q}{\omega^2 +D^2 \lambda_q^2} e^{i q (l-m)}   =\f{1}{2 \pi N} \sum_q \int_{-\infty}^{\infty}  d \omega \f{B^2 \lambda_q}{\omega^2 +D^2 \lambda_q^2} e^{i q(l-m)}  ~, 
\eea
where in the last step we have taken the limit of large $\tau$ to convert the 
summation to an integral. 
This will agree with the equilibrium result if we take
$B^2= 2 D$~.
 
We now evaluate the distribution of the current. In particular, we are interested in the characteristic function $Z(\lambda)=\la e^{-\lambda W} \ra$ 
where the average is over the noise $\widetilde{\eta} (q,\omega)$ and
\begin{align}
W&=\f{1}{N} \sum_{l=1}^N \int_0^\tau dt \phi_l^2 (t) \nn \\
 &=\tau \sum_{q \neq 0} \sum_\omega \widetilde{\phi}(q,\omega) \widetilde{\phi}(-q,-\omega)= \tau \sum_{q \neq 0} \sum_\omega A(q,\omega) \widetilde{\eta}(q,\omega) \widetilde{\eta}(-q,-\omega)~, \nn \\
\text{where}~ ~ A(q,\omega) &=  \f{2 D \lambda_q}{\omega^2+D^2 \lambda_q^2}~. \nn
\end{align} 
Hence we have
\bea
Z(\lambda) = \prod_{q}\prod_{\omega} \la e^{-\lambda \tau A(q,\omega) \widetilde{\eta}(q,\omega) \widetilde{\eta}(-q,-\omega)} \ra=\prod_{q}\prod_{\omega} \f{ N \tau}{ N \tau + \lambda \tau A(q,\omega) }~.
\eea
For large $\tau$ we see that this has the form $Z(\lambda)= e^{\mu(\lambda) \tau}$, where 
\be
\mu(\lambda)= -\f{1}{2 \pi} \sum_{q \neq 0} \int_{-\infty}^\infty d \omega \log \left[ 1+
\f{\lambda}{N} \f{2 D \lambda_q}{\omega^2+D^2 \lambda_q^2} \right]~.
\ee
Using the result $\int_{-\infty}^\infty d \omega \ln [1+ b/(\omega^2+a^2)] =2 \pi [(a^2+b)^{1/2}-a]$, we get
\begin{align}
\mu (\lambda)&=-\sum_{q \neq 0} \left[ \left(D^2 \lambda_q^2 + \f{2D \lambda \lambda_q}{N}\right)^{1/2}-D\lambda_q\right] \nn \\
&= -\sum_{q \neq 0} D \lambda_q \left[ \left(1 + \f{2 \lambda }{D \lambda_q N}\right)^{1/2}-1\right] \nn \\
&= D \sum_{n=1}^\infty \lambda^n ~\f{(-1)^n (2 n) !}{(2n-1) (n!)^2 2^{n} D^n N^n} \sum_{q \neq 0} \f{1}{\lambda_q^{n -1}} ~.
\end{align}
Hence we get
\be
\la W^n \ra_c/\tau =\f{ (2 n) !}{(2n-1) n! 2^{n} D^n N^n} \sum_{q \neq 0} \f{1}{\lambda_q^{n -1}}~.
\ee
Taking the continuum limit, we set $\lambda_q=q^2$ with $q=2 s \pi/N$, and get the following expressions for the even moments of $W$.
\begin{align}
\la W^{2n} \ra_c/\tau &=N^{n-2} \f{ (2 n) !}{(2n-1) (n)! 2^{n} D^{n}} a_n \nn \\
\text{where}~~a_n&=\f{1}{(2 \pi)^{2(n-1)}} \sum_{s=1}^\infty \f{1}{s^{2(n-1)}}= \f{(-1)^n B_{2(n-1)} }{2 (2n-2)!}~, 
\end{align}  
where $B_{2 n}$ are the Bernoulli numbers. 
Hence we get
\be
\la W^{n} \ra_c/\tau = N^{n-2} \f{(-1)^n B_{2(n-1)} }{ (n-1)! (2D)^n }~.
\ee
For the heat current corresponding to Eq.~(\ref{jen3}) we then get for the even cumulants:
\be 
\la Q^{2n} \ra_c/\tau = 2 A^{2n} N^{2n-2} \f{ B_{2(2n-1)} }{ (2n-1)! (2D)^{2n} }~,
\ee
where $A={\kappa \p_\ell P G^0_{11}}/{(c\sqrt{2 \beta})}$. In particular we get
\begin{align}
\la Q^2\ra_c/\tau &= 2 A^2 \f{1/6}{(2D)^2}~, \nn \\
\la Q^4\ra_c/\tau &= 2 A^4 N^2 \f{1/42}{ 3! (2D)^4}~, \nn \\
\la Q^6\ra_c/\tau &= 2 A^6 N^4 \f{5/66}{5!(2D)^6} ~.
\end{align}

\end{document}